\documentclass{appolb}
\usepackage{graphicx}
\usepackage{amsmath}

\begin{document}
\title{Lévy $\alpha$-stable generalization of the ReBB model of elastic proton-proton and proton-antiproton scattering%
\thanks{Presented by I. Szanyi at ``Diffraction and Low-$x$ 2024'', Trabia (Palermo, Italy), September 8-14, 2024.}
}
\author{Tamás Csörgő$^{1,~2~\dag}$, Sándor Hegyi$^{2}$, István Szanyi$^{1,~2,~3~\ddag}$, 
\address{$^1$ELTE E\"otv\"os Loránd University,\\H - 1117 Budapest, P\'azm\'any P. s. 1/A, Hungary;\\
$^2$HUN-REN Wigner RCP, H-1525 Budapest 114, POB 49, Hungary;\\
$^3$MATE Institute of Technology,  KRC,\\H-3200 Gy\"ongy\"os, M\'atrai \'ut 36, Hungary;\\
$^\dag$tcsorgo@cern.ch \\
$^\ddag$iszanyi@cern.ch}
}
\maketitle
\begin{abstract}
The Lévy $\alpha$-stable generalization of the ReBB model of elastic proton-proton and proton-antiproton scattering is presented. The motivation for the future use of this model in describing experimental data is discussed. 
\end{abstract}
  
\section{Introduction}
A. Bialas and A. Bzdak, in 2007, published models for
elastic proton-proton ($pp$) scattering \cite{Bialas:2006qf}, the BB
models for short. In these models, the proton is
described as a bound state of constituent quarks,
and the probability of inelastic $pp$ scattering is
constructed based on R. J. Glauber’s multiple diffractive scattering theory \cite{Glauber:1970jm}. 
In 2015, the BB model was extended: a real part of the scattering amplitude was added in a
unitary \mbox{manner \cite{Nemes:2015iia}}, leading to the Real extended Bialas-Bzdak model, the ReBB model for
short. The $p = (q,d)$ version of the model
that describes the proton as a bound state of a constituent quark and a single-entity constituent diquark is consistent with the
experimentally observed features of elastic $pp$ scattering. The
basic ingredients of the BB model are the inelastic scattering probabilities of two
constituents as a function of their relative transverse position
and the quark-diquark distribution inside the proton. In the BB \cite{Bialas:2006qf} and the ReBB~\cite{Nemes:2015iia} models, the constituent-constituent inelastic scattering probabilities have Gaussian shapes that follow from the Gaussian-shaped parton distributions of the constituents, characterized by the scale parameters $R_q$ and $R_d$. The quark-diquark distribution inside the proton also has a Gaussian shape with scale parameter $R_{qd}$ that characterizes the separation between the quark and the diquark constituents inside the proton. It was found in studies published in 2021 and
2022 \cite{Nemes:2015iia,Csorgo:2020wmw} that the ReBB model describes all
the available data not only on elastic $pp$
scattering, but also on elastic proton-antiproton
($p\bar p$) scattering in a statistically
acceptable manner, \textit{i.e.} with a confidence
level (CL) $\geq$ 0.1\% in the kinematic range,
$0.38 ~{\rm GeV^2} \leq -t \leq 1.2 ~{\rm GeV^2}$ and
$546 ~{\rm GeV} \leq \sqrt{s} \leq 8 ~{\rm TeV}$,
where $t$ is the squared four-momentum
transfer and $\sqrt{s}$ is the squared center of mass
energy. 

\section{Need for an improvement of the ReBB model at low-$|t|$}

\begin{figure}
    \centering
    \includegraphics[width=0.75\linewidth]{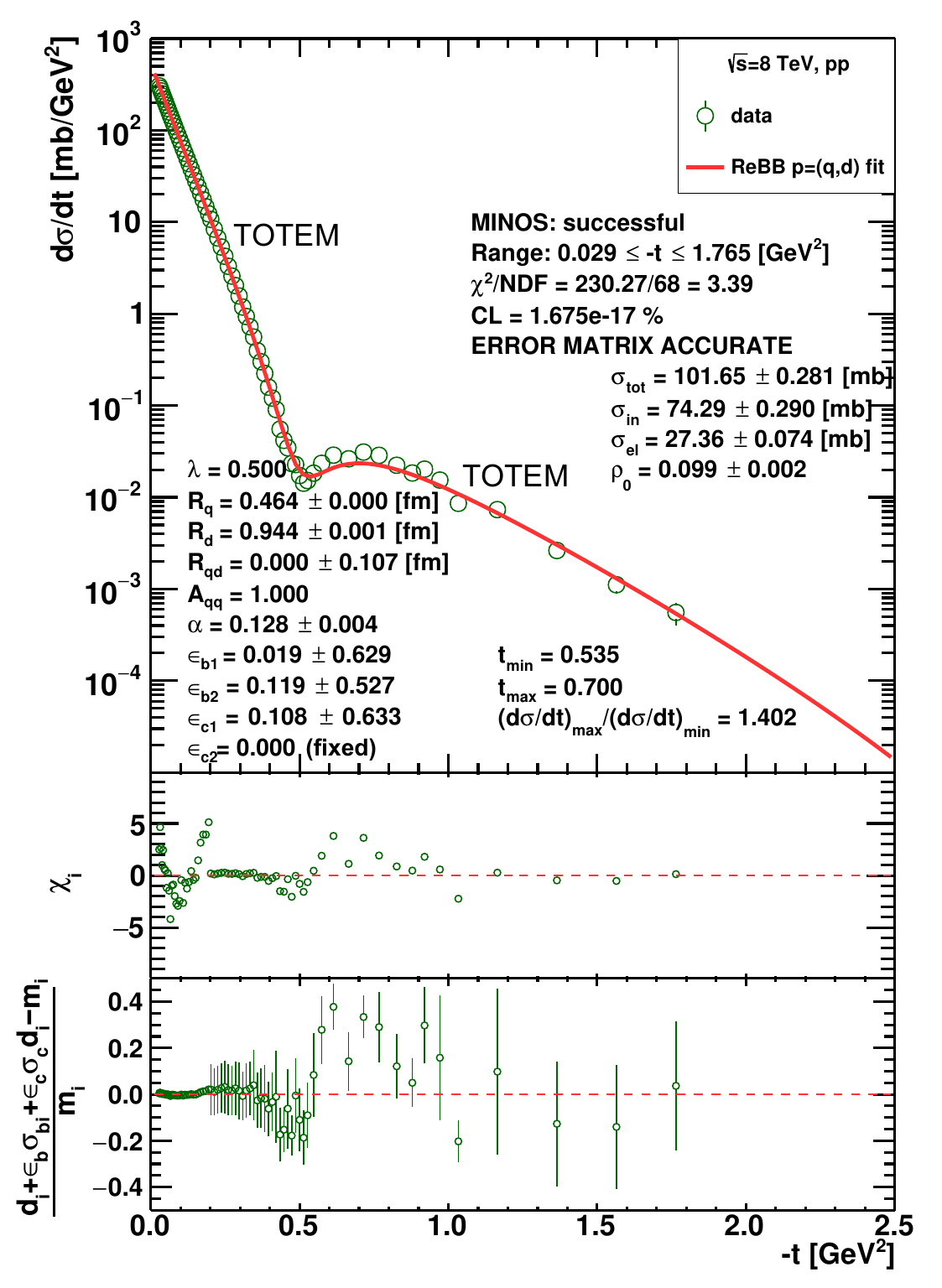}
    \vspace{-3mm}
    \caption{The ReBB model fails to describe simultaneously, with a statistically acceptable confidence level of CL~$> 0.1$\%, both the low-$|t|$ and the high-$|t|$ TOTEM datasets of elastic $pp$ collisions at $\sqrt{s} = 8$ TeV. Similar problems were reported at $\sqrt{s} = 7 $ TeV in Ref.~\cite{Nemes:2015iia}.} 
    \label{fig:RTT}
\end{figure}

\begin{figure}
    \centering
    \includegraphics[width=0.75\linewidth]{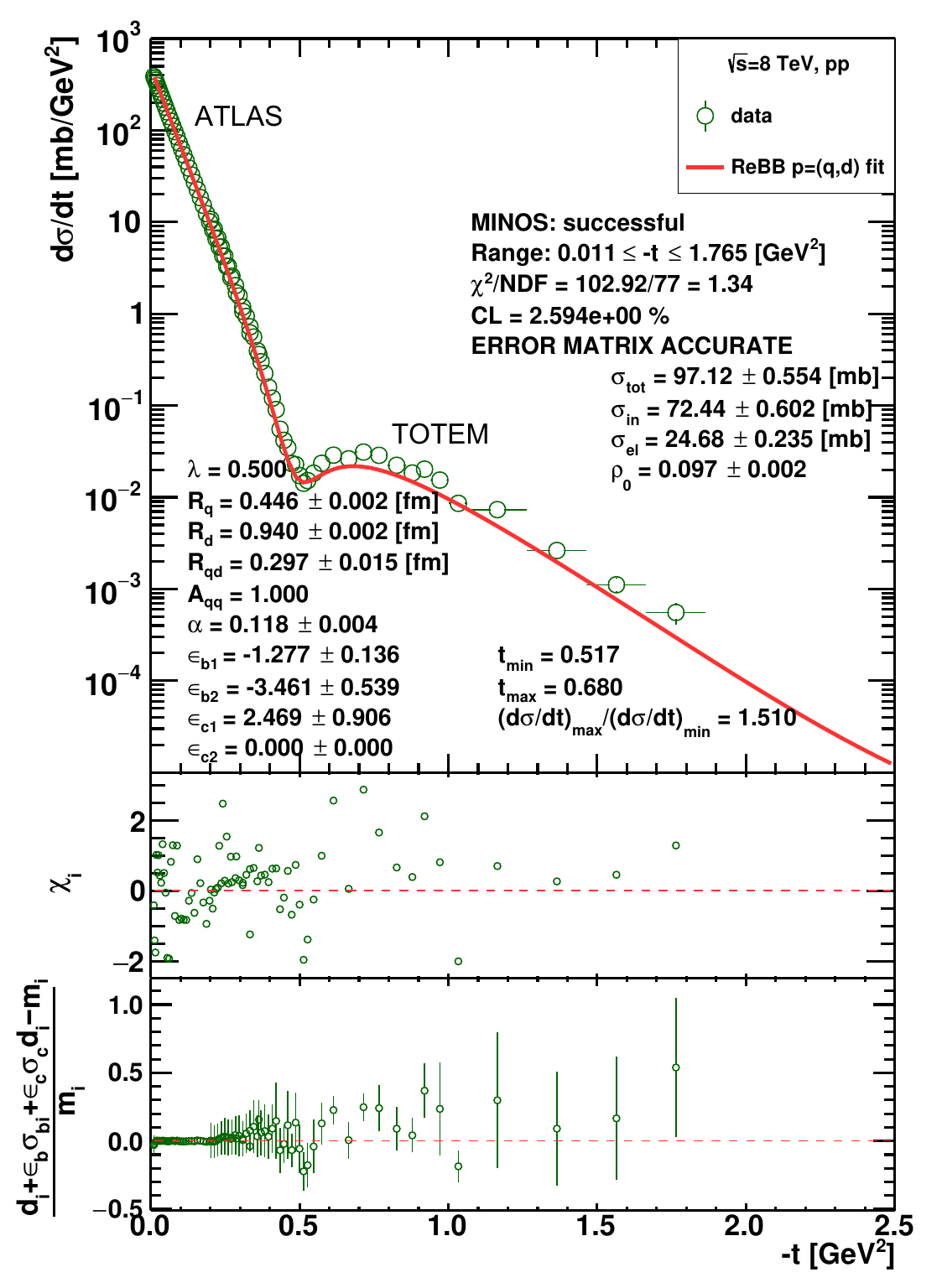}
    \vspace{-2mm}
    \caption{The ReBB model, with an advanced $\chi^2$ definition from the PHENIX experiment \cite{PHENIX:2008ove}, that allows for systematic errors in the slope determination, describes simultaneously, with a statistically acceptable confidence level of CL~$>0.1$\%, the merged low-$|t|$ ATLAS and high-$|t|$ TOTEM datasets of elastic $pp$ collisions at $\sqrt{s} = 8$ TeV. This also resolves the problems reported
at $\sqrt{s} = 7$ TeV in Ref.~\cite{Nemes:2015iia}  when both the low-$|t|$ and the high-$|t|$ TOTEM datasets are described simultaneously, however, the same method fails at $\sqrt{s} = 8$ TeV when both the low-$|t|$ and the high-$|t|$ TOTEM datasets are included (see Fig.~\ref{fig:RTT}).}
    \label{fig:RAT}
\end{figure}

\begin{figure}
    \centering
    \includegraphics[width=0.8\linewidth]{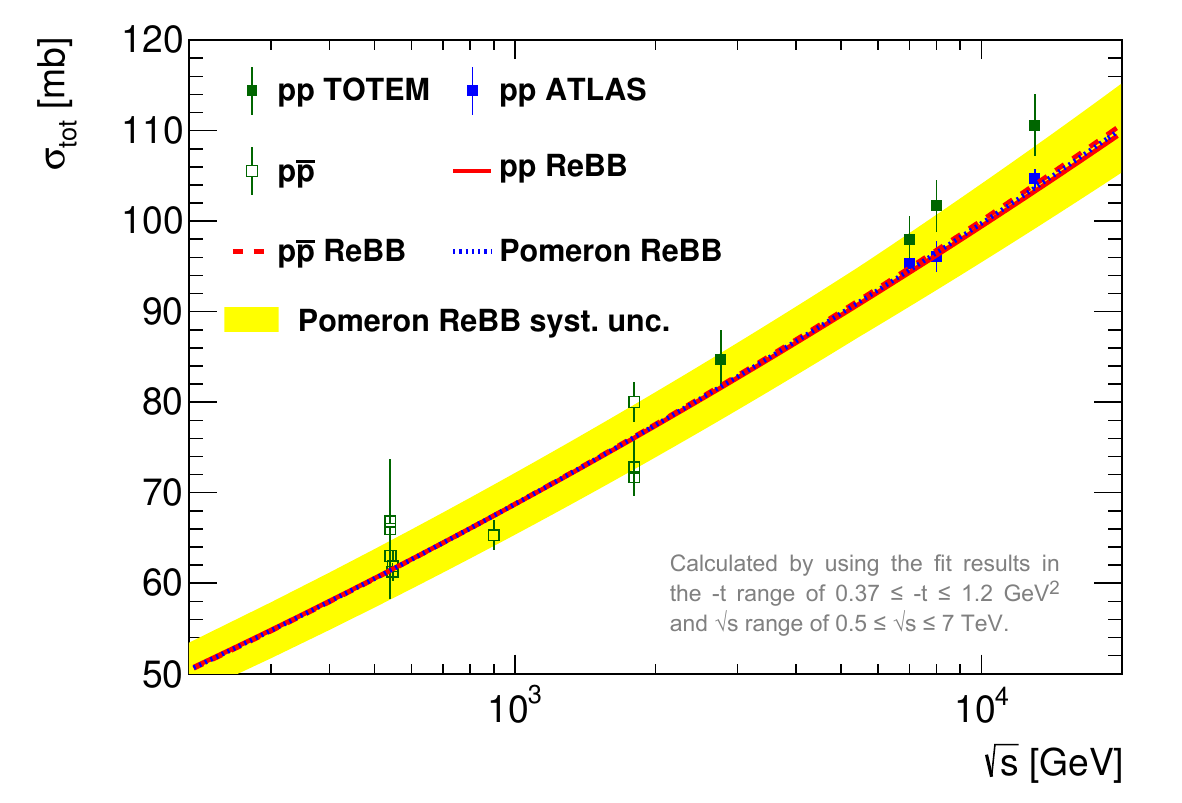}
    \vspace{-2mm}
    \caption{Description to the $\sigma_{\rm tot}$ data by the ReBB model calibrated to the SPS UA4 $p\bar p$,
Tevatron D0 $p\bar p$, and LHC TOTEM $pp$ elastic
$d\sigma/dt$ data in the kinematic range:
$0.38 ~{\rm GeV^2} \leq -t \leq 1.2 ~{\rm GeV^2}$ and 
$546 ~{\rm GeV} \leq \sqrt{s} \leq 7 ~{\rm TeV}$.}
    \label{fig:Rsig_tot}
\end{figure}

At $\sqrt{s}$ = 8 TeV, the ReBB model fails to
describe the TOTEM low-$|t|$ and the TOTEM high-$|t|$ data simultaneously with
CL $\geq$ 0.1\% (see Fig.~\ref{fig:RTT}). The used $\chi^2$ formula is the one
derived by the PHENIX Collaboration \cite{PHENIX:2008ove}. Interestingly, at $\sqrt{s}$ = 8 TeV, the ReBB model describes the ATLAS low-$|t|$ and the TOTEM
high-$|t|$ data simultaneously with CL = 2.6\% (see Fig.~\ref{fig:RAT}).  It is true, however, that the TOTEM low-$|t|$ data shows a strong non-exponential behavior
with a statistical significance greater than 7$\sigma$ \cite{TOTEM:2015oop}, which is not reproduced by the
ReBB model containing Gaussian-shaped distributions. Another interesting feature is that the ReBB model calibrated to the SPS UA4 $p\bar p$,
Tevatron D0 $p\bar p$, and LHC TOTEM pp elastic
$d\sigma/dt$ data in the kinematic range,
$0.38 ~{\rm GeV^2} \leq -t \leq 1.2 ~{\rm GeV^2}$,
\mbox{$546 ~{\rm GeV} \leq \sqrt{s} \leq 7 ~{\rm TeV}$,}
perfectly describes the $pp$ $\sigma_{\rm tot}$ data as
measured by the LHC ATLAS experiment, being
systematically below the $pp$ $\sigma_{\rm tot}$ data as
measured by the LHC TOTEM experiment (see Fig.~\ref{fig:Rsig_tot}).
Theoretically, $\sigma_{\rm tot} (s) = 2{\rm Im}T_{\rm el}(s, t)|_{t\to0}$. Further
studies may be important with a model that is able to describe both ATLAS and TOTEM elastic  $pp$ data both at
low-$|t|$ and high-$|t|$ with CL~$\geq$~0.1\%.

\section{Levy $\alpha$-stable generalized Bialas-Bzdak (LBB) model
}

In the  $p=(q,d)$ BB model the inelastic scattering probability of two protons at a fixed impact parameter vector ($\vec b$) and at fixed constituent transverse position vectors ($\vec s_q$, $\vec s_d$, $\vec s_q^{\,\prime}$, $\vec s_d^{\,\prime}$) is given by a Glauber expansion:
\begin{equation}\label{eq:totalsum}
    \sigma_{\rm in}(\vec b, \vec s_q, \vec s_d, \vec s_q^{\,\prime},\vec s_d^{\,\prime}) = 1- \prod_{a\in\{q,d\}}\prod_{b\in\{q,d\}} \left[1- \sigma_{\rm in}^{ab}(\vec b-\vec s_a + \vec s_b^{\,\prime})\right].
\end{equation}

Generalized central limit theorems
motivate the use of Levy \mbox{$\alpha$-stable}
distributions. In the Lévy $\alpha$-stable generalization of the Bialas-Bzdak (LBB) model, the parton distributions of the constituent quark and the constituent diquark are now Levy $\alpha$-stable distributions, and the inelastic scattering probability for the collision of two constituents at a fixed relative transverse position $\vec x$  of the constituents \cite{Csorgo:2023pdn} is
\begin{align}
         \sigma^{ab}_{\rm in}(\vec s) &= A_{ab} \pi S_{ab}^{2}   \int d^2  s_a L(\vec s_a |\alpha_L,R_a/2) L(\vec s - \vec s_a | \alpha_L, R_b/2 ) \\\nonumber
         &=  A_{ab} \pi S_{ab}^{2} L\left(\vec s\,|\alpha_L, S_{ab}/2\right),
\end{align}

\noindent where $L(\vec x\,|\alpha_L, R_L) = \frac{1}{(2\pi)^2}\int d^2 \vec q e^{i{\vec q}\cdot \vec x}e^{-\left|q^2R_L^2\right|^{\alpha_L/2}}$, $S_{ab}= \left(R_a^{\alpha_L}+R_b^{\alpha_L}\right)^{1/\alpha_L}$, and $a,b\in\{q,d\}$. The distribution of the constituents inside the proton is now given in terms of a Levy $\alpha$-stable distribution ($m_q$ is the mass of the quark and $m_d$ is the mass of the diquark) \cite{Csorgo:2023pdn}:

\begin{equation}\label{eq:quark-diquark_levy}
    D(\vec s_q, \vec s_d) = (1+\lambda)^2 L\left(\vec s_{q} - \vec s_d | \alpha_L, R_{qd}/2\right) \delta^{(2)}(\vec s_d+\lambda \vec s_q),
\end{equation}
where $\int d^2\vec s_q d^2\vec s_d D(\vec s_{q},\vec s_{d}) = 1$, $\lambda=m_q/m_d$, and $\vec s_d=-\lambda  \vec s_q$. 

The probability of inelastic scattering of protons at a fixed $\vec b$ is given by averaging over the constituent positions inside the protons:
\begin{equation}\label{eq:tilde_sigma_inel}
\tilde\sigma_{\rm in}(\vec b)=\int d^{2}\vec s_{q}d^{2}\vec s_{q}^{\,\prime}d^{2}\vec s_{d}d^{2}\vec s_{d}^{\,\prime
}D(\vec s_{q},\vec s_{d})D(\vec s_{q}^{\,\prime},\vec s_{d}^{\,\prime})\sigma_{\rm in}(\vec b, \vec s_q, \vec s_d, \vec s_q^{\,\prime},\vec s_d^{\,\prime}).
\end{equation}

The elastic scattering amplitude $\widetilde T_{\rm el}(s,b)$ is given via  $\tilde\sigma_{\rm in}(s,b)$ -- where $b=|\vec b|$ and the dependence on $s$ follows from the $s$-dependence of the free parameters -- as 
\begin{equation}\label{eq:ReBB_b_ampl}
\widetilde{T}_{\rm el}(s,b)=i\left(1-e^{i\, \alpha_R\, \tilde\sigma_{\rm in}(s,b)}\sqrt{1-\tilde\sigma_{\rm in}(s,b)}\right),
\end{equation}
and  $T_{\rm el}(s,t)$ is obtained via Fourier transformation.

The new free parameter of the Levy  $\alpha$-stable  generalized model is  $\alpha_L$,  the  Lévy index of stability;  if $\alpha_L=2$, the ReBB model with Gaussian distributions is recovered. The power of a simple Lévy $\alpha$-stable model for elastic scattering was demonstrated in Ref. \cite{Csorgo:2023rbs}: good descriptions were obtained to all SPS, Tevatron, and LHC data on low-$|t|$ $pp$ and $p\bar p$ $d\sigma/dt$ in the kinematic range 0.02 GeV$^2\geq |t|\geq$ 0.15 GeV$^2$ and \mbox{546 GeV $\geq\sqrt s \geq$ 13 TeV}, where the Lévy index of stability is compatible with the value of \mbox{1.959 $\pm$ 0.002} implying that the impact parameter distribution has a heavy tail.


\section{Summary}

The Lévy $\alpha$-stable generalization of the Bialas-Bzdak model is made by generalizing from Gaussian shapes to Lévy
$\alpha$-stable shapes both (i) the inelastic scattering probabilities of two constituents and (ii) the quark-diquark
distribution inside the proton. The LBB model is expected to describe simultaneously the low-$|t|$ and high-$|t|$
domains of elastic $pp$ and $p\bar p$ $d\sigma/dt$ with a Lévy index of stability $\alpha <$ 2. Thus the next step is to apply the full LBB
model to describe the data. Given that the LBB model describes the data in a statistically satisfying manner, \textit{i.e.}
with CL~$\geq$~0.1\%, it can be used to study, \textit{e.g.} (i) the discrepancy between ATLAS and TOTEM cross section
measurements, and (ii) after considering the effects of the Coulomb-nuclear interference, the Odderon
contribution to parameter $\rho_0= {\rm Re} T_{\rm el}(s, t)/{\rm Im}T_{\rm el}(s, t)|_{t\to0}$ at $\sqrt{s}$ = 13 TeV.

\section{Acknowledgments}
The research was supported by the ÚNKP-23-3 New National Excellence Program of the Hungarian
Ministry for Innovation and Technology from the source of the National Research, Development and
Innovation Fund; by the NKFIH grant K147557 and 2020-2.2.1-ED-2021-00181; by the Research
Excellence Programme and the Flagship Research Groups Programme of MATE, the Hungarian University of
Agriculture and Life Sciences.

\bibliographystyle{unsrt}  
\bibliography{bib}

\end{document}